\documentclass[aps,prb,twocolumn]{revtex4}
\usepackage{amsfonts}
\usepackage{amsmath}
\usepackage{amssymb}
\usepackage[final,dvips]{graphicx}

\begin{document}

\title{Probing quasiparticle dynamics in Bi$_{2}$Sr$_{2}$CaCu$_{2}$O$_{8 +
\delta }$ with a driven Josephson vortex lattice}
\author{Yu.\ I.\ Latyshev}
\affiliation{Institute of Radio-Engineering and Electronics,
Russian Academy of Sciences, Mokhovaya 11-7, 101999 Moscow,
Russia}
\author{A.\ E.\ Koshelev}
\affiliation{Materials Science Division, Argonne National
Laboratory, Argonne, Illinois 60439}
\author{L.\ N.\ Bulaevskii}
\affiliation{Los Alamos National Laboratory, Los Alamos, NM 87545,
USA} \keywords{} 

\begin{abstract}
We show that the flux-flow transport of the Josephson vortex
lattice (JVL) in layered high-temperature superconductors provides
a convenient probe for both components of quasiparticle
conductivity, $\sigma _{c}$ and $\sigma _{ab}$. We found that the
JVL flux-flow resistivity, $\rho _{ff}$, in a wide range of
magnetic fields is mainly determined by the in-plane dissipation.
In the dense lattice regime ($B>1$ T) $\rho _{ff}(B)$ dependence
is well fitted by the theoretical formula for that limit. That
allows us to independently extract from the experimental data the
values of $\sigma _{c}$ and of the ratio $\sigma _{ab}/(\sigma
_{c}\gamma ^{4})$. The extracted temperature dependence $\sigma
_{ab}(T)$ is consistent with microwave data. The shape of the
current-voltage characteristics is also sensitive to the frequency
dependence of $\sigma _{ab}$ and that allows us to estimate the
quasiparticle relaxation time and relate it to the impurity
bandwidth using data obtained for the same crystal.
\end{abstract}

\maketitle

\section{Introduction}

The properties of quasiparticles (QPs) in the high-temperature
cuprate superconductors are unusual due to d-wave gapless pairing
in these systems. The concentration of QPs in clean d-wave
superconductors vanishes in the limit $T\rightarrow 0$ because
their density of states (DOS) has asymptotics
$\rho(\epsilon)\propto \epsilon$ at low energies,
$\epsilon\rightarrow 0$. Such d-wave features of the QPs spectrum
has been clearly demonstrated by angular-resolved photoemission
(ARPES) studies (see, e.\ g., Refs.\ \onlinecite{ARPES_QP}) and
confirmed by numerous transport measurements. However, QP
transport properties are still in the focus of hot debate and new
experimental techniques are very valuable in understanding of
these properties. In this paper we use the driven Josephson vortex
lattice to probe both in-plane and inter-plane QP transport and
compare results with other techniques.

We start with brief overview of the QP physics in the
high-temperature superconductors. The main issues are how
impurities modify the QPs spectrum and how they affect mobility of
the QPs. Two competing effects determine QP transport at low
temperatures. It was recognized very early that impurities in
unconventional superconductors destroy the superconducting order
parameter in their vicinity increasing the DOS at low
energies.\cite{GorkovKalugin} This effect enhances the
low-temperature QP transport.  On the other hand, increase the QPs
scattering rate by impurities suppresses the QP transport. In
addition, there is a tendency for QPs localization due to the
two-dimensionality of superconducting CuO$_2$ layers.

The low-energy asymptotics of the QP DOS in inhomogeneous d-wave
superconductors is a challenging theoretical problem, which is not
settled yet. Several approximations have been used to calculate
$\rho(\epsilon)$ leading to different results. In the
self-consistent T-matrix approximation (SCTMA)
\cite{LeePRL93,Balatsky,HirschfeldPutikkaScalapino} the DOS
approaches a finite value $\rho(\epsilon)\approx \rho(0)$ for
$\epsilon\lesssim \gamma_i$ and $\rho(\epsilon)\propto \epsilon$
for $\epsilon>\gamma_i$, where the impurity bandwidth $\gamma_i$
depends on the impurity concentration and the impurity potential.
In the limit of strong impurity potential (unitary limit) one gets
for the impurity bandwidth\cite{LeePRL93}
$\gamma_i\approx(\hbar\nu_0\Delta_0)^{1/2}$, where $\nu_0$ is the
normal-state relaxation rate and $\Delta_0$ is the magnitude of
the superconducting gap.

With $\rho(\epsilon)$ known, the QPs transport can be calculated
in the framework of the Fermi liquid theory. Along this line Lee
\cite{LeePRL93} predicted a universal (impurity independent)
low-temperature limit for the intralayer electrical conductivity,
$\sigma_{00}^{(ab)}=ne^2/(\pi m_{ab}\Delta_0)$, arguing that both,
the DOS and the scattering rate are proportional to the impurity
concentration and thus cancel in the Drude expression for the
conductivity. Here $n$ is the carrier concentration and $m_{ab}$
is the QP in-plane effective mass. In the same way universal
thermal conductivity was predicted by Sun and Maki \cite{Maki} and
by Graf {\it et al.}\cite{Graf} Later it was shown in Refs.\
\onlinecite{Fermi} that electron interactions inside the layers
lead to the Fermi-liquid corrections to the universal electrical
conductivity. Durst and Lee \cite{DurstLee}  found that such
corrections as well as the asymmetric scattering (i.e. the
difference between the QPs relaxation rate and the transport
scattering rate) result in the expression for the intralayer
electrical conductivity $\sigma_{ab}(T)= \sigma_{00}^{(ab)}\beta$.
Experimentally observed values of low-temperature $\sigma _{ab}$
usually correspond to $\beta >1$, see Ref.\
\onlinecite{BonnPRL92}. Durst and Lee found also that such
corrections to the universal thermal conductivity at
$\epsilon\lesssim \gamma_i$ are practically absent.

For the interlayer tunneling conductivity, $\sigma_c$, in highly
anisotropic layered cuprates like BSCCO the authors of
Ref.~\onlinecite{LatPRL99} have argued that it is universal in the
limit of low impurity concentration, when electrons tunnel between
layers conserving their in-plane quasi-momentum (coherently). In
contrast to $\sigma_{ab}$, $\sigma_c$ depends only on the
intralayer DOS (i.e., on the QPs relaxation rate). It is not
sensitive to  anisotropy of the in-plane scattering and in-plane
vertex corrections are not important.
In this case the low-temperature conductivity $\sigma_{00}^{(c)}$
is related to the Josephson critical current density $J_c(0)$ by a
simple relation similar to the well-known Ambegaokar-Baratoff
relation in conventional Josephson junctions,
$\sigma_{00}^{(c)}=2esJ_c(0)/(\pi \Delta_0)$, where $s$ is the
interlayer spacing.
In the framework of SCTMA-Fermi-liquid approach the temperature
corrections to the universal conductivity $\sigma_{00}^{(c)}$  at
$T<\gamma_i$ were found \cite{LatPRL99} to be
\begin{equation}
\sigma_c(T)\approx \sigma_{00}^{(c)}[1+(\pi T)^2/18\gamma_i^2].
\label{cT}
\end{equation}
A similar result has been obtained for the in-plane QP
conductivity neglecting anisotropic scattering and vertex
corrections.\cite{HirschfeldPutikkaScalapino} However, it is known
that such corrections substantially modify $\sigma_{ab}(T)$. For
the thermal conductivity, $\kappa(T)/T$, in the unitary limit, the
thermal corrections are also quadratic in $T/\gamma_i$, as for
$\sigma_c$, and only in the limited temperature interval linear in
$T$ corrections were found if the impurity potential is not very
strong.\cite{HirschfeldPutikka}

The more elaborate approaches, which take into account
interference effects, suggest that the QP DOS vanishes at
$\epsilon \rightarrow 0$ (see recent review in Ref.\
\onlinecite{AltlandPR02}). In particular, recent numerical
analysis \cite{AtkinsonHirschfeld} have demonstrated that QP DOS
behaves at very low energies as $\rho(\epsilon)\propto
\epsilon^{\alpha}$ with a nonuniversal exponent $\alpha$. This
exponent depends on the details of disorder model and
particle-hole symmetry in the normal state. In the realistic case
(binary alloy model without particle-hole symmetry) they found
$\alpha>0$, i.e., a DOS suppression at low energies. The energy
scale for this suppression is given is the resonance energy for an
isolated impurity $\Omega_0$ with $\Omega_0\rightarrow 0$ as the
impurity potential increases. This means that at very low
temperatures, $T\ll\Omega_0$, QP transport should vanish.

Intralayer conductivity was studied by a microwave technique
\cite{BonnPRL92,SigmaQBSCCO,Trunin,CorsonPRL00} and infrared
spectroscopy.\cite{RomeroPRL92,BasovPRL95,SantanderSyroPRL02}
These measurements have shown that $\sigma_{ab}(T)$ is not
universal in the low temperature limit and that its temperature
dependence at $T\sim \gamma_i$ deviates strongly from SCTMA
predictions,\cite{HirschfeldPutikkaScalapino} see below. The
impurity bandwidth can be estimated from the frequency dependence
of the intralayer conductivity, $\sigma _{ab}(\omega )$. According
to the model calculation of Ref.\
\onlinecite{HirschfeldPutikkaScalapino}, the typical relaxation
rate, $1/\tau$, in $\sigma _{ab}(\omega )$ has to be of the order
of $\gamma_i$. Recent terahertz spectroscopy measurements of
$\sigma _{ab}(\omega )$ in Bi$_{2}$Sr$_{2}$CaCu$_{2}$O$_{8 +
\delta }$ (BSCCO) by Corson \textit{et al.} \cite{CorsonPRL00}
showed that at low temperatures it has a Drude frequency
dependence with the typical relaxation rate $1/\tau \approx 1$
THz. This gives an estimate $\gamma_i= 30-50$K.

Experimentally a universal value of $\kappa_{00}/T$ was confirmed
by measurements \cite{KrishanaPRL95,Movsh,Nakamae} of the thermal
conductivity in BSCCO and YBCO crystals. Particularly, Nakamae
{\it et al.} \cite{Nakamae} measured thermal conductivity in
pristine and irradiated BSCCO crystals at $T<0.1$ K. Direct
comparison of $\kappa(T)$ in pristine and irradiated crystals
showed almost the same value $\kappa_{00}/T$ obtained by
extrapolation to $T=0$. However, strong temperature corrections,
linear in $T$, to this universal value were found below 0.25 K.
Such corrections are not anticipated well below $\gamma_i\approx
20$ K in the SCTMA approach. One can conclude from these
measurements, that the upper limit for the low temperature
behavior, $\Omega_0$, lies below 0.1 K and that temperature
behavior of $\kappa(T)$ observed so far is in contradiction with
the theoretical SCTMA predictions.

Measurements of interlayer conductivity seems to follow closely to
the theoretical predictions of the the SCTMA-Fermi-liquid model
assuming coherent interlayer tunneling.\cite{LatPRL99}  For BSCCO
crystals from different groups, similar values,
$\sigma_{00}^{(c)}\approx 2$ [k$\Omega\cdot$ cm]$^{-1}$ were
observed by measurements of the $I$-$V$ characteristics, though
universality of this value was not checked by comparison of
crystals before and after irradiation as it was done for the
thermal conductivity.\cite{Nakamae} The temperature dependence of
$\sigma_c(T)$ at $T<\gamma_i$ was found to be in agreement with
the theoretical prediction, Eq.\ (\ref{cT}), and the values of
$\gamma_i$ in the interval 24-29 K were derived from fitting, in
agreement with the data for the relaxation rate $\tau$ mentioned
above.

The intralayer and interlayer components of the QP conductivity
are found to have qualitatively different temperature dependencies
in the superconducting state. The interlayer conductivity
monotonically decreases with temperature decrease in the whole
temperature range from $T_c$ down to lowest temperatures. In
contrast, the intralayer conductivity has manifestly nonmonotonic
behavior: it rapidly \emph{increases} with decrease of temperature
in some region below $T_c$ reaches a peak at some intermediate
temperature, and decreases with further decrease of temperature.
\cite{BonnPRL92,SigmaQBSCCO,Trunin,CorsonPRL00} This behavior is
also consistent with the heat transport measurements for the
electronic part of the thermal conductivity.\cite{KrishanaPRL95}
The peak in the temperature dependence of $\sigma _{ab}$ appears
due to an interplay between the temperature dependencies of the
relaxation rate \cite{CorsonPRL00} and concentration of
quasiparticles. The unusual increase of the QP conductivity and
thermal conductivity just below $T_c$ is attributed to the fast
drop of the relaxation rate due to the reduction of the phase
space for scattering.\cite{HirschfeldPutikka} Decrease of the
conductivity at low temperatures, below the peak, is caused by a
drop in the concentration of the thermally activated nodal
quasiparticles.

Hence, the behavior of QPs in the cuprate d-wave superconductors
is not fully understood yet and new methods to probe the QP
transport would be useful to resolve controversy and provide
additional information on the characteristic parameters, such as
$\Omega_0$.

\begin{figure}[tbp]
\includegraphics[clip,
width=3.2in ]{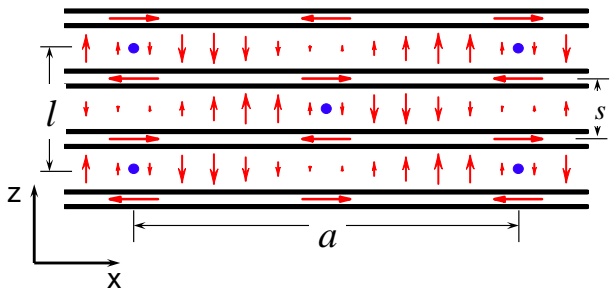} \caption{Dense Josephson vortex
lattice with period $l$ = 2$s$ in the $c$-direction. Arrows show
the direction of currents.} \label{Fig-DenseLat}
\end{figure}

Recently a new method of probing the QP conductivities $\sigma_c$
and $\sigma_{ab}$ has been suggested.\cite{KoshPRB00} The method
is based on the measuring of the losses, associated with a
transport of the Josephson vortex lattice (JVL) \cite{DenseJosLat}
driven by steady current across the layers in crystals with
intrinsic Josephson interlayer coupling.\cite{BulPRB96} The
unexpected dependence of the interlayer transport on the
intralayer quasiparticle conductivity in the flux-flow regime is
related to the spatially inhomogeneous structure of moving JVL.
Figure \ref{Fig-DenseLat} shows the space structure of a
stationary high-field Josephson lattice.\cite{DenseJosLat} Inside
the layers the supercurrent oscillates along the direction
perpendicular to the applied parallel magnetic field ($x$ -axis).
The interlayer transport current drives the vortex lattice along
the $x$-axes. At small velocities the lattice is practically
undistorted. Then the supercurrent inside the layer at a given
point changes periodically with time, $j_{sx}\propto $ $\sin (2\pi
vt/a)$, where $v$ is the lattice velocity and $a=\Phi _{0}/sH$ is
the period of vortex lattice. According to the first London
equation (see, e.g., Ref.\ \onlinecite{Tinkham})
\begin{equation*}
E_{x}=\frac{4 \pi \lambda _{ab}^{2}}{c^2}\frac{\partial
j_{sx}}{\partial t},\quad \lambda_{ab}^{2}=\frac{m_{ab}c^{2}}{4\pi
n_{s}e^{2}},
\end{equation*}
an alternating electric field $E_{x}$ with frequency $\omega =2\pi
v/a$ is introduced by a moving vortex lattice. Here $n_{s}$ is the
density of superconducting electrons. The electric field causes
relaxation due to quasiparticle current $j_{qx}=\sigma
_{ab}E_{x}$. Hence, as was shown, both components of the QP
conductivity $\sigma _{c}$ and $\sigma _{ab}$ contribute to the
Josephson flux-flow resistivity $\rho _{ff}$ and can be extracted
separately from the magnetic field dependence of $\rho _{ff}$.
Moreover, the shape of the I-V characteristic related to JVL
motion at high magnetic fields is sensitive to the frequency
dependence of the QP conductivity and therefore may be used to
estimate the QP relaxation time. This approach allows one to probe
the relatively low frequency range of the QP conductivity, 0.01-3
THz, and thus the results can be easily compared with the data of
microwave measurements. The goal of the present paper is to
demonstrate the applicability of this new method for studies of
the QP conductivity in BSCCO.

Motion of JVL induced by a steady current across the layers
results in a specific branch on the I-V characteristic usually
referred to as the Josephson flux-flow (JFF) branch. That is
characterized by a rapid current increase when the voltage
approaches a certain limiting value, $V_{m}$, at which the lattice
velocity approximately reaches the velocity of electromagnetic
wave propagation (this velocity frequently is refered to as the
Swihart velocity in analogy with a single long Josephson
junction). The JFF regime is well known for conventional long
Josephson junctions \cite{Barone} and has also been observed on
BSCCO mesa structures.\cite{LeeAPL95,HechPRL97,LatyshPhysC97} Our
purpose was to study the JFF linear and non-linear regimes on long
BSCCO stacks at high magnetic fields above 0.5 T when a dense JVL
is formed. Early experiments on JFF in BSCCO have been done mostly
at relatively low fields \cite{LeeAPL95,LatyshPhysC97} or at high
fields of only few fixed values.\cite{HechPRL97}
In addition, an absence of the clear upturn curvature on JFF
branch \cite{HechPRL97} suggests significant inhomogeneities in
the mesas used in early experiments.

We report new measurements that allow us to obtain for the first
time the parameters $1/\tau$ and $\gamma_i$ for the same crystal
and correlate the frequency and temperature dependencies of
quasiparticle conductivities.
\begin{figure}[tbp]
\includegraphics[clip,width=3.4in ]{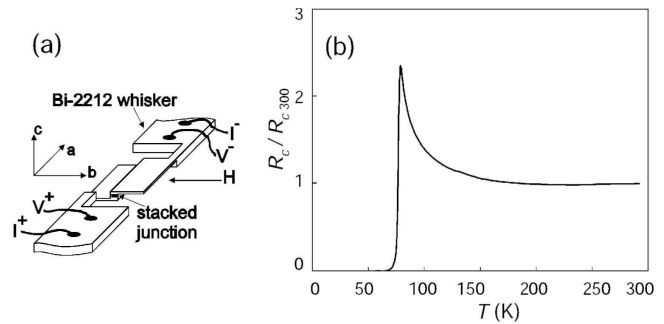} \caption{(a)Schematic view of
junction orientation and experimental set up.
(b)Normalized temperature dependence of resistance along the
c-axis of typical long stacked junction used in experiment.}
\label{Fig-SchemView}
\end{figure}

\section{Experimental}

To obtain high-quality stacks we fabricated samples from single
crystal whiskers of BSCCO. The thin BSCCO whiskers have been
characterized as extremely perfect crystalline objects
\cite{LatPhysC93}. They grow along the [100] direction free of any
crucible or substrate and can be entirely free of macroscopic
defects and dislocations. The stacks have been fabricated by the
double-sided processing of the BSCCO whiskers by the focused ion
beam (FIB) technique. The stages of fabrications were similar to
those described in Ref.\ \onlinecite{LatIEEE99}. Figure
\ref{Fig-SchemView}a shows schematically the geometry and
orientation of the structure with respect to the crystallographic
axes. We reproduced the overlap type of long stack geometry which
is known to provide the most uniform current distribution along
the junction.\cite{Barone} The structure sizes were $L_{a}=20-30$
$\mu $m, $L_{b}=1-2$ $\mu$m, $L_{c}=0.05-0.15$ $\mu$m.
Using high-resolution optical microscope we selected for the
experiment long uniform whiskers with a length of 500-1000 $\mu
$m, a thickness of 0.5-1 $\mu $m and a width of 10-20 $\mu $m.
Four silver contact pads have been evaporated and annealed at
450$^\circ$C in oxygen flow before FIB processing to avoid
diffusion of Ga-ions into the junction body. The fabricated
structures have been then tested by $R_{c}(T)$ measurements to
select ones free of inclusions of 2201 or 2223 phases. The
presence of these phases is indicated as a multiple transition to
superconducting state with appropriate drops of $R_{c}(T)$ at
90-100K for 2223 phase and at 15-30K for 2201 phase. Only
single-phase 2212 stacked junctions with a single transition at
75-80 K (see Fig.\ \ref{Fig-SchemView}b) have been selected for
the further measurements. A yield of the single-phase stacks was
quite high, about 30-50{\%}.
The oxygen doping level of the stacks estimated from $\rho
_{c}(T)$ measurements above $T_{c}$ \cite{WatanPRL97} was slightly
above optimum, $\delta \approx 0.25$. The critical current density
$J_{c}$ at 4.2 K in the absence of magnetic field was 1-2
kA/cm$^{2}$. Measurements of the I-V characteristics of BSCCO
stacks have been carried out in the commercial cryostat of Quantum
Design PPMS facility. The magnetic field has been oriented
parallel to the $b$-axis within accuracy $0.1^{\circ }$ and has
been changed in steps of 0.05-0.1 T. In each fixed value of the
field the I-V characteristics have been measured using a fast
oscilloscope. We have measured 6 samples with similar results.
\begin{figure*}[tbp]
\includegraphics[clip,width=6in ]{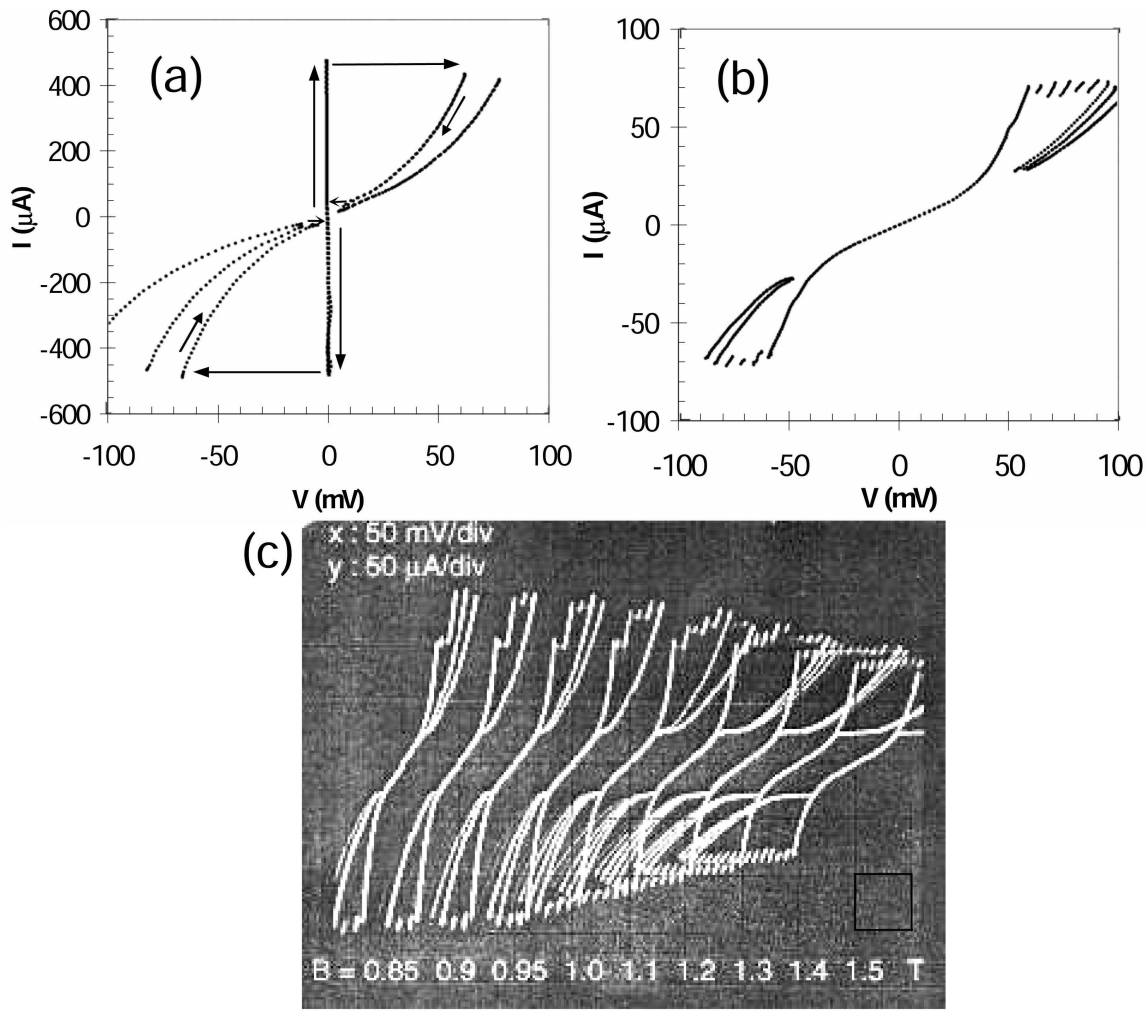}
\caption{The I-V characteristics of stack {\#}4 at 4.2 K without
magnetic field (a) and with field $B$ applied along $b$-axis,
$B=1.5$T (b), with $B$ increasing from 0.85 T up to 1.5 T (c). The
first three branches are not traced on the I-V characteristic in
zero field and the stack jumps directly to the 4th  branch.}
\label{Fig-ExperimIv}
\end{figure*}

\section{RESULTS AND DISCUSSION}

Fig.\ \ref{Fig-ExperimIv}  shows the I-V characteristics of a long
stack \#4 ($30\times 2\times 0.14$ $\mu$m$^3$) at $T = 20$ K at
zero magnetic field (Fig.\ \ref{Fig-ExperimIv}a) and at the field
$B=1.5$T (Fig.\ \ref{Fig-ExperimIv}b) oriented along the b-axis.
At zero magnetic field the I-V characteristics of the stacks show
a well-defined critical current shown at Fig.\
\ref{Fig-ExperimIv}a as a vertical trace with the following switch
to the multibranched structure. The I-V characteristic is highly
hysteretic. The hysteresis loop is shown in Fig.\
\ref{Fig-ExperimIv}a schematically by arrows (the first switch
here corresponds to the jump from the critical current to the
4$^{th}$ branch). Both features, the hysteresis and the
multibranched structures of the I-V characteristics, are well
known for Bi-2212 stacked junctions at low temperatures.
\cite{KleinPRL92}

\begin{figure}[tbp]
\includegraphics[clip, width=3.2in ]{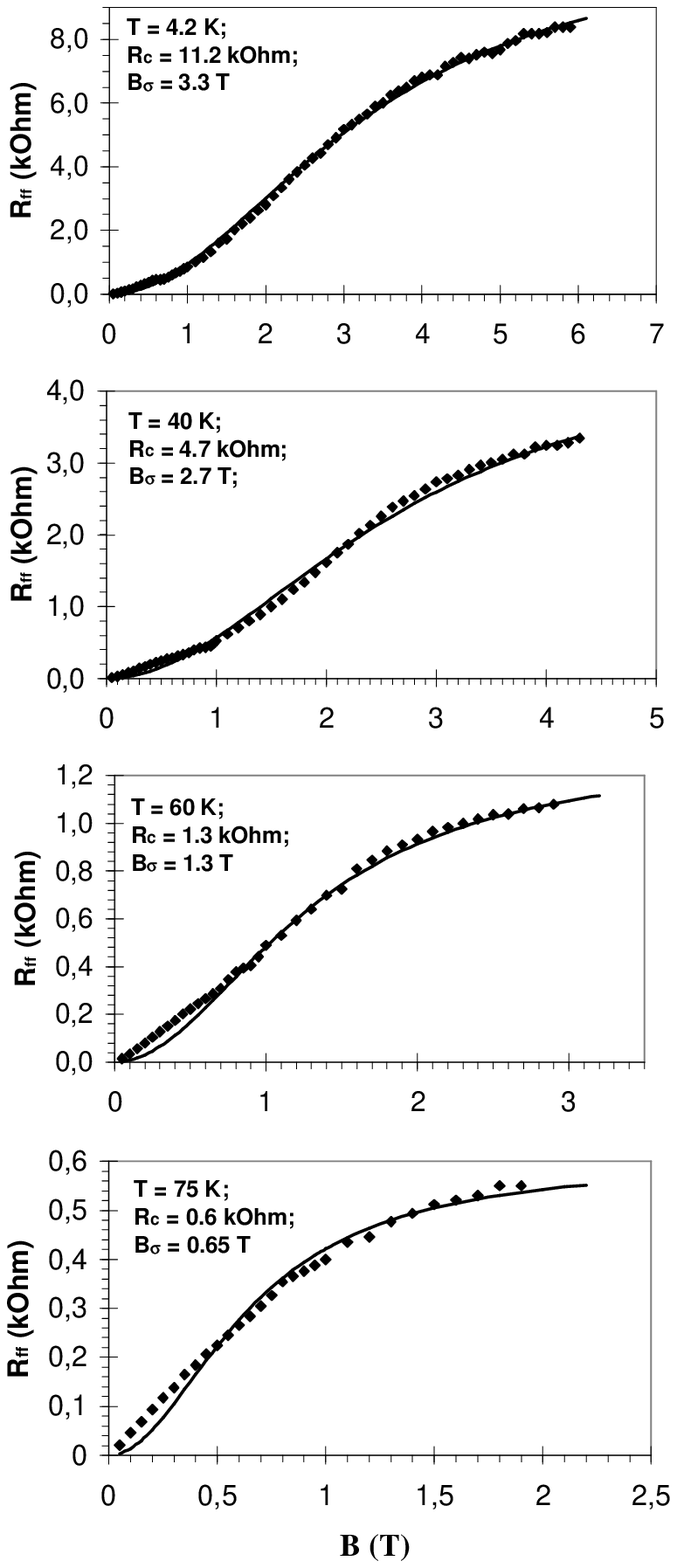}
\caption{Magnetic field dependence of the Josephson flux-flow
resistance, $R_{ff}$ at different temperatures with fits to Eq.\
(\ref{rhoffHigh}).} \label{Fig-Rho_ff}
\end{figure}

\begin{figure}[tbp]
\includegraphics[clip, width=3.2in ]{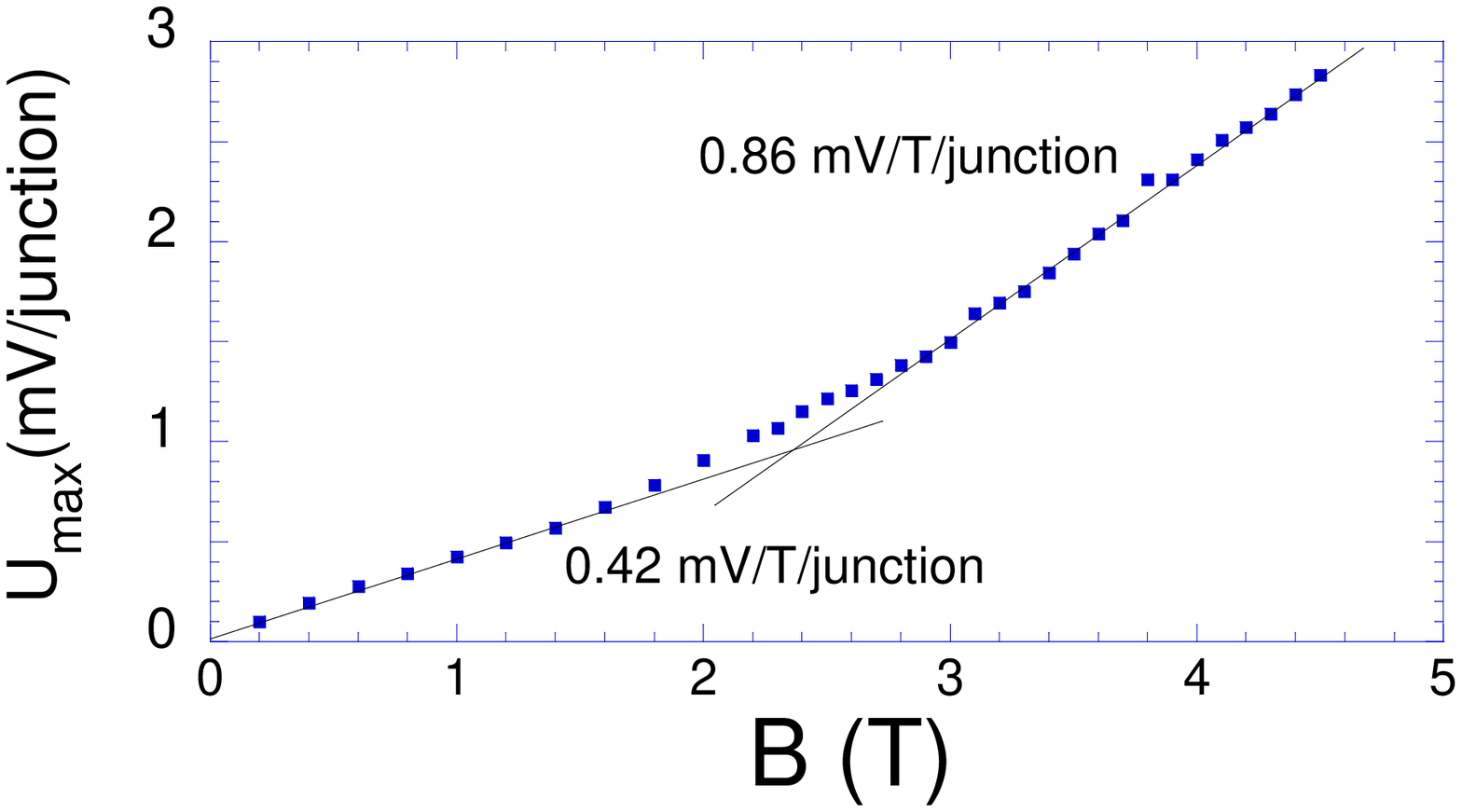}
\caption{Magnetic field dependence of maximum voltage of Josephson
flux-flow branch, $V_{max}$.} \label{Fig-VmaxB}
\end{figure}

In the presence of a parallel field the critical current becomes
essentially suppressed and the JFF branch develops. That is
characterized by a linear slope at low bias, $R_{ff}$, by a
pronounced upturn at higher bias voltages, and by jumps to the
multiple branches at the voltages exceeding the maximum value
$V_{m}$. The high upturn in the nonlinear I-V characteristics in
the parallel field proves a high quality of our stacks. To get the
$R_{ff}(B)$ dependence we measured  a set of the I-V
characteristics at some temperatures for the fixed fields
increasing step by step.
At each field we measured $R_{ff}$. The value of $R_{ff}$ has been
defined as an extrapolation of the linear part of the I-V at
$\vert V\vert  \to 0$. As seen from Fig.\ \ref{Fig-ExperimIv}c,
this extrapolation can be easily done at high fields when critical
current is highly suppressed. An accuracy of linear extrapolation
of the I-V characteristic at $\vert V\vert  < 10$mV for definition
of $R_{ff}$ was within 5{\%}.
Fig.\ \ref{Fig-ExperimIv}c shows an evolution of the I-Vs with
field. One can see a rapid increase of $R_{ff}$ and $V_{m}$ with
field. The summarized field dependencies of $R_{ff}$ and $V_{m}$
for typical sample ({\#}4) are shown at Figs.\ \ref{Fig-Rho_ff}
and \ref{Fig-VmaxB}. Before analyzing these data we discuss the
expected theoretical dependencies.

Firstly, we will consider the linear limit of the I-V
characteristics corresponding to low JFF velocity. In the second
part, we will focus on the high-velocity JFF  limit.

\subsection{Linear flux-flow resistivity.}

The linear flux-flow resistivity of the Josephson vortex lattice,
$\rho _{ff}$, is determined by the static lattice structure and
linear quasiparticle dissipation. At high fields, $B>\Phi _{0}(\pi
\gamma s^{2})$, the Josephson vortices homogeneously fill all the
layers and the static lattice structure is characterized by
oscillating patterns of both $c$-axis and in-plane supercurrents
\cite{DenseJosLat} (see Fig.\ \ref{Fig-DenseLat}). Here $\gamma$ is
the anisotropy ratio of the London penetration lengths,
$\gamma=\lambda_c/\lambda_{ab}$. At small
velocities this pattern slowly drifts along the direction of
layers, preserving its static structure. This motion produces
oscillating $c$-axis ($\tilde{E}_{z}$) and in-plane
($\tilde{E}_{x}$) electric fields leading to extra dissipation, in
addition to usual dissipation due to the \textit{dc} electric
field $E_z$. Total dissipation per unit volume is given by
\begin{equation}
\sigma _{ff}E_{z}^{2}=\sigma _{c}E_{z}^{2}+\sigma _{c}\left\langle
\tilde{E} _{z}^{2}\right\rangle +\sigma _{ab}\left\langle
\tilde{E}_{x}^{2}\right\rangle,  \label{sigmaff}
\end{equation}
where $\sigma _{ff}=1/\rho _{ff}$ is the flux-flow conductivity, $
\left\langle \ldots \right\rangle $ means time and space average,
$\sigma _{c}=1/\rho _{c}$ and $\sigma _{ab}=1/\rho _{ab}$ are the
c-axis and in-plane quasiparticle conductivities. An expansion
with respect to the Josephson current at high fields allows to
relate $\tilde{E}_{z}$ and $\tilde{E}_{x}$ with $E_{z}$,
\begin{eqnarray*}
\tilde{E}_{nz}(x,t) &=&-(-1)^{n}\frac{4E_{z}}{h^{2}}\cos \left(
k_{H}x+\omega t\right), \\
\tilde{E}_{nx}(x,t) &=&(-1)^{n}\frac{2E_{z}}{\gamma h}\sin \left(
k_{H}x+\omega t\right),
\end{eqnarray*}
with $h=2\pi \gamma s^{2}B/\Phi _{0}$ and $k_H=2\pi sB/\Phi_0$.
Here $n$ is the layer index. Finally, we obtain a simple
analytical formula for the flux-flow resistivity
\cite{KoshPRB00,rhoff_footnote}
\begin{equation}
\rho _{ff}(B)=\frac{B^{2}}{B^{2}+B_{\sigma }^{2}}\rho
_{c},\;\;B_{\sigma }=\sqrt{\frac{\sigma _{ab}}{\sigma
_{c}}}\frac{\Phi _{0}}{\sqrt{2}\pi \gamma ^{2}s^{2}},
\label{rhoffHigh}
\end{equation}
The relative importance of the in-plane and $c$-axis dissipation
channels is determined by the dimensionless ratio $\Gamma =\sigma
_{ab}/(\sigma _{c}\ \gamma ^{2})$. The in-plane dissipation
channel dominates when $\Gamma \gg 1$.

Experimental dependencies of the JFF resistance, $R_{ff}(B)$, on
magnetic field $B$ applied along the $b$-axis at various
temperatures are depicted in Fig.\ \ref{Fig-Rho_ff}. They are
evidently nonlinear with the type of non-linearity predicted by
Eq.\ (\ref{rhoffHigh}): quadratic growth with the following
saturation. As shown in Fig.\ \ref{Fig-Rho_ff}, the data can be
well fitted to Eq.\ (\ref{rhoffHigh}) in the wide range of
temperatures. Two fitting parameters have been used for each
curve, $R_{c}$, the high field saturation value of the resistance,
$R_{ff}(B)$, and the characteristic field $B_{\sigma }$, both
defined by Eq.\ (\ref{rhoffHigh}). Using that fit at a set of
temperatures we found a temperature variation of both parameters.
Then, from the temperature dependence of $R_{c}$ we can directly
get the dependence of $\sigma _{c}(T)$, while the temperature
dependence $B_{\sigma }$ contains information about the ratio
$\sigma _{ab}(T)/\sigma _{c}(T)$ for given values of $\gamma $.
For T = 4.2 K we found $B_{\sigma }=3.3$ T. We estimate the value
of $\gamma \approx 500$ at 4.2 K from the value of the Josephson
critical current density $J_c(0)=1.7$ kA/cm$^{2}$ and the value of
$\lambda _{ab}(0)=0.2$ $\mu $m using the well-known expression
\cite{LawDon}, $J_c=c\Phi _{0}/(8\pi ^{2}s\gamma ^{2}\lambda
_{ab}^{2})$. That gives a quite reasonable value for $\sigma
_{ab}$ at 4.2 K, $\sigma _{ab}(4.2)=4\cdot 10^{4}$ (Ohm
cm)$^{-1}$. Finally, we restored the temperature dependence of the
in-plane quasiparticle conductivity $\sigma _{ab}(T)$ taking into
account the temperature dependence of $\gamma $. We used the
$\gamma (T)$ dependence extracted from the known data for $\lambda
_{ab}(T)$ \cite{JacobsPRL95} and $\lambda _{c}(T)$
\cite{GaifPRL99} in BSCCO. Actually, $\gamma ^{2}$ only weakly
depends on $T$, slowly decreasing within 15\%, with temperature
increase from 4.2 to 70 K.
\begin{figure}[tbp]
\includegraphics[clip,
width=3.in ]{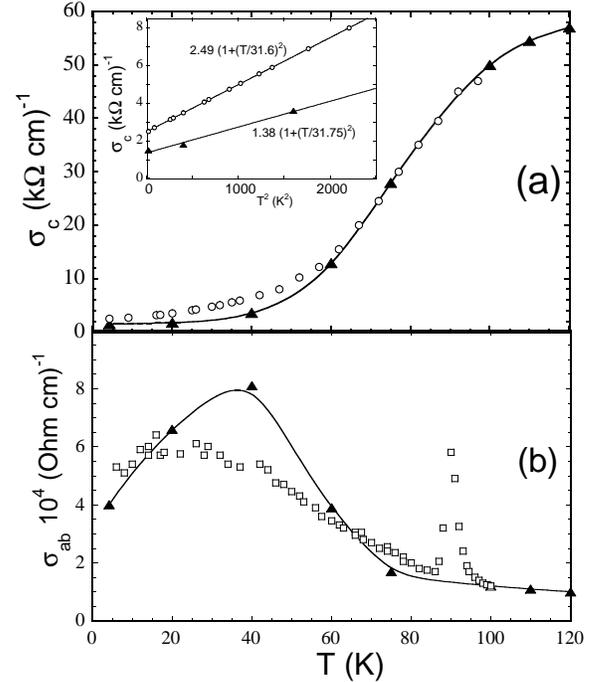} \caption{Solid triangles show
temperature dependencies of the out of plane quasiparticle
conductivity $\sigma _{c}$ (a) and in-plane quasiparticle
conductivity $\sigma_{ab}$ (b). Below T$_c$ they extracted from
the JFF experiment on BSCCO long stack {\#}4  and  above $T_c$
they represent the \textit{dc} normal-state conductivities of
whiskers measured independently on samples from the same batch.
Open circles correspond to the $\sigma _{c}$ data from Ref.\
\onlinecite{LatPRL99}, obtained on small mesas in zero field, open
squares correspond to 14.4 GHz microwave data for $\sigma_{ab}$
from Ref.\ \onlinecite{SigmaQBSCCO} obtained on BSCCO epitaxial
films.  Solid lines in both plots are just guides to the eye.
Inset in the upper figure shows the low-temperature part of
$\sigma_c(T)$ plotted versus $T^2$. } \label{Fig-sigmas}
\end{figure}

Figure \ref{Fig-sigmas} demonstrates the temperature dependencies
of $\sigma _{ab}$ and $\sigma _{c}$ extracted from the Josephson
flux-flow experiment. As we found, $\sigma _{ab}$ rapidly
increases below $T_{c}$ with decrease of $T$, reaches a maximum at
about 30K, and then drops down at low temperatures. That type of
$\sigma _{ab}(T)$ behavior has been found earlier in the microwave
experiments for YBCO\cite{BulPRB96} and BSCCO
\cite{SigmaQBSCCO,Trunin,CorsonPRL00} and also is consistent with
the heat transport measurements of the electronic part of the
thermal conductivity.\cite{KrishanaPRL95} As discussed in the
Introduction, the peak in the temperature dependence of $\sigma
_{ab}$ appears due to an interplay between the competing
temperature dependencies of the relaxation rate \cite{CorsonPRL00}
and concentration of quasiparticles. We found that the value and
temperature dependence of $\sigma _{ab}$ extracted from the
Josephson flux-flow experiment quite well reproduce the
low-frequency microwave data.\cite{SigmaQBSCCO} The sharper
increase of $\sigma _{ab}(T)$ below $T_{c}$ may be related to
weaker scattering of the quasiparticles in the BSCCO
single-crystal whiskers used in our experiments. At higher
temperatures
the data are also consistent with the results of DC measurements
of $\sigma _{ab}(T)$ in the normal state carried out at the
whiskers from the same batch \cite{LatPhysC93} (see the points at
$T\geq 100$ K). We see that, in contradiction with the naive SCTMA
predictions, \cite{HirschfeldPutikkaScalapino}
$\sigma_{ab}(T)/\sigma_{ab}(0)-1 \sim (T/\gamma_i)^2$, $\sigma
_{ab}(T)$  has strong temperature dependence at $T<\gamma_i\approx
30$ K so that the saturation to the low-temperature value is not
reached even at 4 K.

In contrast, the temperature dependence of $\sigma _{c}$ extracted
from the flux-flow experiment (Fig.\ \ref{Fig-sigmas}a) is
consistent with $\sigma _{c}(T)$ measured on small mesas in zero
magnetic field,\cite{LatPRL99} with $\sigma _{c}(0)$ being close
to the universal value $\sigma _{c}(0)\quad \approx 2$ (k$\Omega $
cm)$^{-1}$ predicted in Ref.\ \onlinecite{LatPRL99}. The data also
well extrapolate to the points of \textit{dc} measurements of
$\sigma _{c}(T)$ of the same mesas in the normal state. The
temperature dependence of $\sigma_c(T)$ is in agreement with the
theoretical prediction Eq.~(\ref{cT}) with $\gamma_i\approx 23.5$
K.

\begin{figure}[tbp]
\includegraphics[clip,
width=3.in ]{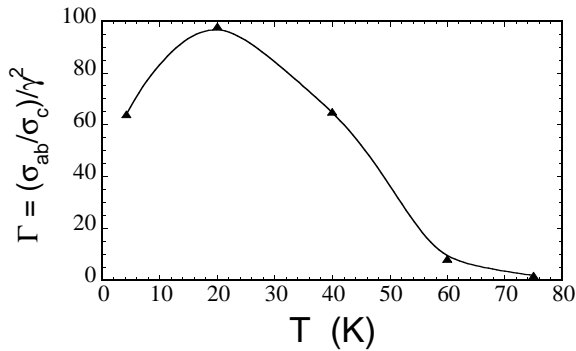} \caption{ Temperature dependence of the
parameter $\Gamma =(\sigma _{ab}\sigma _{c})/\gamma ^{2}$. Solid
line is a guide to eyes.} \label{Fig-Gamma_T}
\end{figure}

Clearly, behavior of $\sigma_{ab}(T)$ and $\sigma_c(T)$ is very
different. This can be seen more clear in the temperature
dependence of their ratio. In Fig.\ \ref{Fig-Gamma_T} we plot
temperature dependence of the parameter $\Gamma =(\sigma
_{ab}/\sigma _{c})/\gamma ^{2}$ (the ratio of the QP dissipation
anisotropy to the superconducting anisotropy). The important point
is that both components of the quasiparticle conductivity were
extracted from the same experiment. As was mentioned above the
in-plane contribution to the flux-flow resistivity becomes
considerable when $\Gamma \gg 1$. Figure \ref{Fig-Gamma_T} shows
that this condition is indeed valid in a wide temperature range
below $T_{c}$. Note that the value of $\Gamma $ approaches 1 when
$T$ approaches $T_{c}$. The value of $\Gamma $ also decreases
below 20 K. The interesting issue is the low-temperature limit of
$\Gamma $. Without Fermi-liquid and anisotropic scattering
corrections to the universal electrical intralayer conductivity,
$\Gamma$ should approach unity at low temperatures,
$T\ll\gamma_i$, but obtained $\Gamma$ is still well above unity at
4 K.

\subsection{Nonlinear Josephson flux-flow regime.}

A non-linear Josephson flux-flow occurs at high velocities of
Josephson vortex lattice, especially at velocities approaching the
minimum velocity of the electromagnetic wave (Swihart velocity),
$c_s=cs/(2\lambda_{ab}\sqrt{\epsilon})$.
The main source of the nonlinearity is the pumping the energy from
a \textit{dc} source into the travelling electromagnetic wave,
generated by the moving lattice. Due to the deformation of the the
moving lattice induced by interaction with the
boundaries,\cite{KoshAran01} the I-V characteristics in this
regime can be calculated only numerically. Below we present the
main steps of a calculation (see Ref.\ \onlinecite{KoshAran01} for
details) as well as a comparison with the experiment.
We will consider here only the main flux-flow branch corresponding
to the case of uniformly sliding triangular lattice. The
termination point of this branch is related to instability of the
moving triangular vortex lattice.\cite{KoshAran01,ArtRemPRB03} The
high-voltage branches in the JFF regime are apparently related
with more complicated behavior and beyond the scope of this paper.
We emphasize that at high velocity, when the washboard frequency
exceeds the inverse quasiparticle relaxation time, one has to take
into account the frequency dependence of the quasiparticle
conductivity \cite{CorsonPRL00} which leads to the renormalization
of the plasma frequency and Swihart velocity.

At high fields in the resistive state the interlayer phase
differences depend approximately linearly on coordinate and time
\begin{equation}
\theta _{n}(t,x)\approx \omega_{E}t+k_{H}x+\phi _{n},
\label{Phase0}
\end{equation}
where $\omega _{E}$ is the Josephson frequency and $k_{H}$ is
magnetic wave vector. In the following we will use reduced
parameters: $\omega _{E}\rightarrow \omega _{E}/\omega _{p}$,
$k_{H}\rightarrow k_{H}\gamma s$ (see Table 1). The most important
degrees of freedom in this state are the phase shifts $\phi _{n}$,
which describe the structure of the moving Josephson vortex
lattice. In particular, for the static triangular lattice $\phi
_{n}=\pi n$. The lattice structure experiences a nontrivial
evolution with increase of velocity. The equations for $\phi _{n}$
can be derived from the coupled sine-Gordon equations for $\theta
_{n}(t,x)$ by expansion with respect to the Josephson current and
averaging out fast degrees of freedom.\cite{KoshAran01} In the
case of a steady state for a stack consisting of $N$ junctions,
this gives the following set of equations
\begin{equation}
\frac{1}{2}\sum_{m=1}^{N}\mathrm{Im}\left[ \mathcal{G}(n,m)\exp
\left( i\left( \phi _{m}-\phi _{n}\right) \right) \right] =i_{J},
\label{PhaseDyn}
\end{equation}
where
$n=1,2\ldots N$
and $i_{J}\equiv i_{J}(k_{H},\omega
_{E})=\left\langle \sin \theta _{n}(t,x)\right\rangle $ is the
reduced Josephson current, which has to be obtained as a solution
of these equations. The complex function $\mathcal{G}(n,m)$
describes phase oscillations in the $m$-th layer excited by the
oscillating Josephson current in the $n$-th layer. For a finite
system it consists of the bulk term $G(n-m)$ plus the top and
bottom reflections (multiple reflections can be neglected):
\begin{equation*}
\mathcal{G}(n,m)=G(n-m)+\mathcal{B}G(n+m)+\mathcal{B}G(2N+2-n-m),
\end{equation*}
where
\begin{eqnarray}
G(n) &=&\int \frac{dq}{2\pi }\frac{\exp \left( iqn\right) }{\omega
^{2}-i\nu
_{c}\omega -\Omega ^{2}(k,q,\omega )},  \notag \\
\Omega ^{2}(k,q,\omega ) &=&\frac{k^{2}\left( 1+i\nu _{ab}\omega
\right) }{2(1-\cos q)+\left( 1+i\nu _{ab}\omega \right) /l^{2}}
\label{Omega_kq}
\end{eqnarray}
$\omega =\omega _{E}$ and $k=k_{H}$ are the frequency and the
in-plane wave vector of the travelling electromagnetic wave
generated by moving lattice.
The real and imaginary parts of $\Omega ^{2}(k,q,\omega )$ give
the spectrum of the collective plasma oscillations and their
dumping due to in-plane quasiparticle dissipation (in reduced
units).
The dissipation parameters, $\nu _{c}$ and $\nu _{ab}$, and
reduced penetration depth $l$ are defined in Table 1.
$\mathcal{B}=\mathcal{B}(k,\omega )$ is the amplitude of reflected
electromagnetic wave. For the practical case of the boundary
between the static and moving Josephson lattices a detailed
calculation of $\mathcal{B}(k,\omega )$ is presented in Ref.\
\onlinecite{KoshAran01}.

In general, the quasiparticle conductivities in the definitions of
$\nu _{c}$ and $\nu _{ab}$ are the complex conductivities at the
Josephson frequency. The frequency dependence is especially
important for the in-plane conductivity. The maximum Josephson
frequency at the termination point of the flux-flow branch exceeds
the value $1/\tau$ at fields $B\gtrsim 2$ T. Therefore the
frequency dependence has to be taken into account. We use the
Drude-like frequency dependence of $\nu _{ab}\equiv \nu
_{ab}(\omega )$:
\begin{equation}
\nu _{ab}(\omega )=\frac{\nu _{ab0}}{1+i\omega \tau }
\label{nuab_omega}
\end{equation}
where $\tau $ is the quasiparticle relaxation time. The solution
of Eq.\ (\ref{PhaseDyn}) yields the current-voltage characteristic
\[
j(E_{z})=\sigma_{c}E_{z}+J_c\,i_{J}(k_{H},\omega _{E}),
\]
where $k_{H}$ and $\omega _{E}$ has to be expressed via magnetic
and electric fields (see Table 1).

\begin{figure}[tbp]
\includegraphics[clip, width=3.in ]{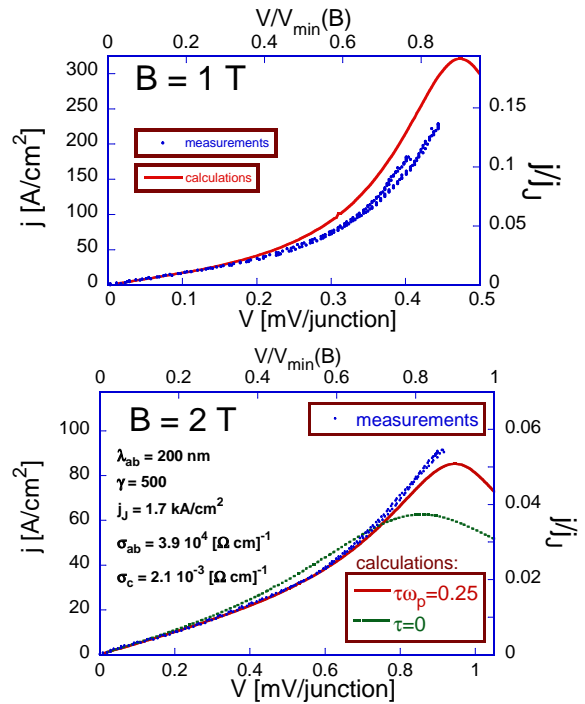}
\caption{Comparison of the experimental I-V characteristics
measured on long BSCCO stack at fields $B = 1$T and $B = 2$T at
4.2 K with calculated numerically using Eq.\ (\ref{PhaseDyn}). For
both cases $\gamma $ was used as a fitting parameter. The best fit
was found at $\gamma = 500$. For the I-V characteristic at B=2 T
also $\tau $ has been varied. The best fit was found at $\omega
\tau  = 0.25$. For such relaxation rate frequency dependence of
$\sigma _{ab}$ have very weak influence on the I-V shape at 1 T.
In the upper axes the voltage scale $V_{\min}(B)\equiv
Bs^2/(2\lambda_{ab}\sqrt{\epsilon})\sim 0.5$ mV/T corresponds to
the voltage at which the lattice velocity reaches the minimum
velocity of electromagnetic wave
$sc/(2\lambda_{ab}\sqrt{\epsilon})$. The parameters assumed in
calculations are listed in the lower plot.} \label{Fig-IV_comp}
\end{figure}

We solved Eqs.\ (\ref{PhaseDyn}) numerically and calculated the
I-V dependencies for the first flux-flow branch. We used $\sigma
_{c}$ and $\sigma _{ab}/\gamma ^{4}$ obtained from the fit of
$\rho _{ff}(B)$, assumed $\lambda _{ab}=200$ nm, and adjusted
$\gamma$ to obtain the I-V dependencies most close to experimental
ones. The results are shown in Fig.\ \ref{Fig-IV_comp} for two
values of magnetic field, $B=1$ T and $B=2$ T at $T=4.2$ K. One
can see a very reasonable fit to experiment for both field values
using $\gamma \approx 500$. At high fields (see curves at 2 T) we
found the fit can be significantly improved by taking into account
frequency dependence of $\sigma _{ab}$ via Eq.\
(\ref{nuab_omega}). The best approximation here was found for
$1/(2\pi \tau )=0.6$ THz.  Both fitting parameters found to be
quite reasonable. The value for $\gamma $ is consistent with the
typical value of $J_c(0)$ for our samples at low temperatures,
$J_c(0)=1-2$ A/cm$^{2}$, while the value of $1/(2\pi \tau )$ is
consistent with the microwave data of Corson \textit{et al.}
\cite{CorsonPRL00} where the relaxation rate at low temperature
was found to be $\approx 1$ THz.

\subsection{Renormalization of Swihart velocity by quasiparticles}

Relatively low quasiparticle relaxation rates lead to an important
observable consequence: renormalization by quasiparticles of the
Swihart velocity at high frequencies, $\omega >1/\tau $.
The Swihart velocity is the in-plane velocity of the
electromagnetic wave at the maximum c-axis wave vector,
$k_z=\pi/s$.
For this mode, out-of-phase oscillations of c-axis current in the
neighboring layers induce oscillations of the in-plane current
leading to strong coupling with the in-plane charge transport. As
a consequence, the  velocity of this mode,
$cs/(2\lambda_{ab}\sqrt{\epsilon})$, is strongly reduced by the
Cooper pairs in comparison with velocity of the transverse
electromagnetic wave, $c/\sqrt{\epsilon}$. At small frequencies
the in-plane motion of quasiparticles only contributes to the
dissipation. However, when frequency exceeds the scattering rate,
quasiparticles contribute to the inductive response in the same
way as the superconducting electrons do, reducing an effective
screening length and increasing the Swihart velocity. We will
analyze this effect quantitatively.
The spectrum of the plasma mode, $\omega _{p}=\omega _{p}(k,q)$
and its damping parameter $\nu =\nu (k,q)$ are determined by the
equation (reduced units) $\omega _{p}^{2}+i\nu \omega _{p}=i\nu
_{c}\omega _{p}+\Omega ^{2}(k,q,\omega _{p})$ with $\Omega
(k,q,\omega _{p})$ defined by Eq.\ (\ref{Omega_kq}), i.e., $\omega
_{p}(k,q)$ is given by the solution of the equation
\begin{equation*}
\omega _{p}^{2}=\mathrm{Re}[\Omega ^{2}(k,q,\omega _{p})]
\end{equation*}
For the minimum frequency at fixed $k$ corresponding to $q=\pi $
we obtain
\begin{equation*}
\omega _{p}^{2}\approx \frac{k^{2}}{4}\left( 1+\mathrm{Re}\left[
\frac{\nu _{ab}i\omega _{p}}{1+i\tau \omega _{p}}\right] \right)
\end{equation*}
From this equation one can easily observe that the Swihart
velocity $c_{s}=\omega _{p}/k$ (in units of
$sc/(\lambda_{ab}\sqrt{\epsilon})$) has two simple asymptotics
\begin{equation*}
c_s=\genfrac{\{}{.}{0pt}{}{1/2\text{, at
}c_{s}k\tau\ll1}{\sqrt{1+\nu_{ab}/\tau }/2\text{, at
}c_{s}k\tau\gg1}
\end{equation*}
The physical meaning of the quasiparticle renormalization factor
$r=\sqrt{1+\nu _{ab}/\tau }$ becomes more transparent if we
transfer to the real units and use the two-fluid expressions for
$\lambda _{ab}$ and $\sigma _{ab}$, $\lambda
_{ab}^{2}=m_{ab}c^{2}/(4\pi n_{s}e^{2}),\;\sigma _{ab}=n_{n}\tau
e^{2}/m_{ab}$. Then the renormalization factor reduces to $\nu
_{ab}/\tau =n_{n}/n_{s}$ and
\begin{equation*}
c_s=\genfrac{\{}{.}{0pt}{}{\sqrt{\pi s^{2}n_{s}e^{2}/\varepsilon
m_{ab}}\text{, at }c_{s}k\tau\ll1}{\sqrt{\pi s^{2}\left(
n_{s}+n_{n}\right)  e^{2}/\varepsilon m_{ab}}\text{, at
}c_{s}k\tau\gg1}
\end{equation*}
i.e., the renormalization amounts to replacement of the superfluid
density $n_{s}$ by the total density $n_{s}+n_{n}$.

The first flux-flow branch terminates due to instability of the
triangular lattice configuration at the velocity close to the
Swihart velocity.\cite{KoshAran01,ArtRemPRB03} Therefore, the
Swihart velocity can be estimated from the voltage at the endpoint
of the first flux-flow branch.
As a consequence of quasiparticle renormalization, one can expect
an increase of the maximum voltage for the first flux-flow branch
when the Josephson frequency at this voltage exceeds $1/\tau $.
The estimate of the renormalization factor for our samples for
$\omega _{p}\tau =0.25$, $\lambda _{ab}=0.2$ $\mu $m, $\sigma
_{ab} = 4\cdot 10^{4}$ ($\Omega $ cm)$^{-1}$ gives the value $r =
2$. Experimentally, the Swihart velocity can be extracted from the
slope of the linear dependence of the maximum flux-flow voltage on
the magnetic field, $V_{ffmax}(H)/H$. We found (see also Ref.\
\onlinecite{HechPRL97}), that the slope increases at magnetic
fields above 1.5 T (Fig.\ \ref{Fig-VmaxB}). That field corresponds
to the washboard frequency about 0.5 THz, which is very close to
the quasiparticle relaxation rate. Therefore, the increase of the
slope of $V_{ffmax}(H)$ at higher frequencies may be interpreted
as an increase of the Swihart velocity due to a renormalization of
plasma frequency at $\omega >1/\tau $. Experimentally, the slope
of $V_{ffmax}(H)$ increases by factor 2.1 that is very close to
the theoretical estimate of the renormalization factor.

\section{Conclusions}

In summary, we carried out detailed experimental and theoretical
studies of the linear and non-linear Josephson flux-flow regimes
in BSCCO.  Both regimes as shown can be well described by the
developed theoretical model. That takes into account an additional
channel in the dissipation related to the \textit{ac} in-plane
quasiparticle currents accompanying JFF. The criterion of the
applicability of the model $\sigma _{ab}/(\sigma _{c}\ \gamma
^{2})>1$ is shown to be valid in a wide temperature range below
$T_{c}$. From the measurements in the linear JFF regime we
extracted the values of both components of the quasiparticle
conductivity at low frequency limit. The extracted temperature
dependence $\sigma _{ab}(T)$ is consistent with the microwave
measurements, while $\sigma _{c}(T)$ reproduces the \textit{dc}
measurements on small mesas in zero magnetic field. The fit of the
data measured in the nonlinear regime to the theoretical model
allowed us to estimate quasiparticle relaxation time $\tau $ and
the renormalized Swihart velocity at high frequency limit $\omega
\tau
> 1$. All these results demonstrate the applicability of JFF
measurements for studies of quasiparticle dynamics in layered
high-$T_{c}$ superconductors.

We derived from the experimental data for $\sigma_c(T)$ the
impurity bandwidth $\gamma_i$ and from the nonlinear part of the
$I$-$V$ characteristics we estimated the QPs relaxation rate
$1/\tau$, which is consistent with $\gamma_i$.
In a similar way, the  universality of $\sigma_{00}^{(c)}$ may be
checked by measurements of pristine and irradiated crystals and
upper limit for $\Omega_0$ may be given by study of $\sigma_c(T)$
at lower temperatures and frequencies.

\section{Acknowledgements}

We would like to thank N.\ Pedersen and A.\ Ustinov, for fruitful
discussions, M.\ Graf and K.\ Gray for critical reading of the
manuscript and constructive comments, V.\ N.\ Pavlenko for
technical assistance, and T.\ Yamashita for support of this work.
YIL acknowledges support from the CRDF grant No.\ RP-12397-MO-02
and grant from Russian Ministry of Science and Industry No.\
40.012.1.111.46. In Argonne this work was supported by the U.\ S.\
DOE, Office of Science, under contract No.\ W-31-109-ENG-38. AEK
and YIL acknowledge support from the NATO Travel grant No.\
PST.CLG.979047.

\newpage
\begin{widetext}
{\bf Table 1.} Meanings, definitions and practical formulas for
the reduced parameters used in the paper. In practical formulas
$f_{p}=\omega _{p}/2\pi $ means plasma frequency, $\rho _{c}$ and
$\rho _{ab}$ are the components of the quasiparticle
resistivity\newline
\begin{tabular}[t]{|c|c|c|c|}
\hline Notation & Meaning & Definition(CGS) & Practical formula
(BSCCO) \\ \hline $\omega _{E}$ & {\small reduced Josephson
frequency} & {\Large {$\frac{2\pi
csE_{z}}{\Phi _{0}\omega _{p}}$}} & {\Large {$\frac{U[{\rm mV/junction}]}{%
2\cdot 10^{-3}f_{p}[{\rm GHz}]}$}} \\ \hline
$k_{H}$ & {\small magnetic wave vector} & {\Large {$\frac{2\pi H\gamma s^{2}%
}{\Phi _{0}}$}} &  \\ \hline $\nu _{c}$ & {\small c-axis
dissipation parameter} & {\Large {$\frac{4\pi \sigma
_{c}}{\varepsilon _{c}\omega _{p}}$}} & {\Large {$\frac{1.8\cdot
10^{3}}{\varepsilon _{c}\rho _{c}[\Omega \cdot {\rm cm}]f_{p}[{\rm
GHz}]}$}}
\\ \hline
$\nu _{ab}$ & {\small in-plane dissipation parameter} & {\Large
{$\frac{4\pi
\sigma _{ab}\lambda _{ab}^{2}\omega _{p}}{c^{2}}$}} & {\Large {$\frac{%
0.79(\lambda _{ab}[\mu m])^{2}f_{p}[{\rm GHz}]}{\rho _{ab}[\mu
\Omega \cdot {\rm cm}]}$}} \\ \hline
$l$ & {\small reduced London penetration depth} & $\lambda _{ab}/s$ &  \\
\hline
\end{tabular}

\end{widetext}



\begin{thebibliography}{99}
\bibitem{rhoff_footnote} The same result was obtained earlier in Ref.\ \onlinecite{BulPRB96},
Eq.\ (42), but with a misprint in the sign in front of the term
with $\sigma _{ab}$, where plus should stay.
\end{thebibliography}
\end{document}